\documentclass[twocolumn,aps,prd,amsmath,amssymb,superscriptaddress]{revtex4}
\usepackage{bm}%
\input epsf
\begin{document}

\title{\bf Classical Gravitation as free Membrane Dynamics
}
\author{{Miguel D. Bustamante}}
\email{mig_busta@yahoo.com}
\affiliation{Laboratoire de Physique
Statistique de l'Ecole Normale
Sup\'erieure, associ\'e au CNRS et aux
Universit\'es Paris VI et VII, 24 Rue
Lhomond, 75231 Paris, France}
\affiliation{{Mathematics Institute,
University of Warwick, Coventry CV4 7AL,
United Kingdom}}
\author{{Fabrice Debbasch}}
\email{debbasch.fabrice@wanadoo.fr} \affiliation{{ERGA, UMR 8112, 4
Place Jussieu, F-75231 Paris Cedex 05, France}}
\author{{Marc-Etienne Brachet}}
\email{marc_brachet@noos.fr} \affiliation{Laboratoire de Physique
Statistique de l'Ecole Normale Sup\'erieure, associ\'e au CNRS et
aux Universit\'es Paris VI et VII, 24 Rue Lhomond, 75231 Paris,
France}

\begin{abstract}

The formulation of General Relativity in
which the $4$-dimensional space-time is
embedded in a flat host space of higher
dimension is reconsidered. New classes of
embeddings (modeled after Nash's classical
free embeddings) are introduced. They
present the important advantage of being
deformable and therefore physically
realistic. Explicit examples of embeddings
whose deformations \emph{do} describe
gravitational waves around their respective
backgrounds are given for several
space-times, including the Schwarzschild
black hole. New variational principles
which give back Einstein's General
Relativity  are proposed. In this
framework, the 4-D space-time is a membrane
moving in a flat host space.

\vspace{0.3cm}
\par\noindent
{\bf PACS numbers\,:04.20.Cv, 04.20.Fy, 04.30.Nk}
\par\noindent
{\bf Keywords\,:} General Relativity; Embeddings; Variational
Formulations; Waves.
\end{abstract}
\maketitle

\section{Introduction}
\label{sec:Intr}

General Relativity is commonly regarded as the correct approach to
non-quantum gravitation \cite{W95a}. Einstein's theory views gravity
as a manifestation of the curvature of the $4$-D space-time
\cite{W84a}. Several authors have proposed to consider this physical
curved $4$-D space-time as a membrane embedded in a flat space-time
of higher dimension called the host space \cite{Reg75,
Des76,Pav85,{Mai89},Tap89,Fra92,Pav94,bandos-1997-12,{Pav01}}. This
point of view is computationally convenient and is also extremely
natural in the context of modern string and brane theory
\cite{randall-1999-83,Mai02,maartens-2004-7,kiritsis-2005-421,davidson-2006-74}.
The aim of the present article is to complement the existing
literature on this topic. Our main conclusion is that the embedding
approach to GR can be successfully implemented in a
large variety of contexts and provides some undeniable
computational and conceptual advantages.
Here follows a summary of our
principal results.

We first introduce two new classes of embeddings (modeled after
Nash's classical free embeddings  \cite{Nas56}) and explain why
these two classes are particularly natural from the physical point
of view. Although they typically require host spaces of higher
dimensions than most embeddings proposed by various authors
\cite{Fro59,Ros65,lidsey-1997-14,deser-1998-15,kim-2000-62}, these
new classes of embeddings present the important physical advantage
of being deformable, and therefore physically more realistic. In
particular, given an arbitrary space-time, any embedding of this
space-time which belongs to one of the two new classes can be
deformed to obtain an embedding for gravitational waves propagating
in this space-time.

We then give explicit examples of embeddings in both classes for the
standard Minkovski space-time, the Schwarzschild black hole and
gravitational waves propagating in flat space-time. We then propose
new variational principles which give back Einstein's General
Relativity by viewing the 4-D space-time as a membrane moving in a
flat host space. Some of the variational principles involve new
border terms previously not considered by previous authors.
Actually, the issue of constructing actions which deliver the equations of standard
General Relativity in terms of embedding functions has been often
addressed in the literature \cite{{Reg75},{Des76},{Tap89},{Fra92}}.
Our work is the first to propose a solution to this long standing
problem. We finally show that the embedding
point of view permits a particularly simple and physically
enlightening treatment of the initial value problem in relativistic
gravitation.

\section{Free embeddings}
\label{sec:Isometrical v/s Free}

\subsection{Generalities about embeddings}
\subsubsection{What is an embedding?}
We denote the physical 4-D space-time by $\mathcal M$ and its
Lorenztian, possibly curved metric by $g$. Space-time indices
running from $0$ ($1$) to $3$ will be indicated by Greek (Latin)
letters and the metric signature will be $(+, -, -, -)$. The
covariant derivative for tensor fields defined on $\mathcal M$ is,
as usual, the derivative operator associated with the Levi-Civita
connection of the metric $g$. We also consider a `host'-space
${\mathcal{E}}_N$ \emph{i.e.} an $N$-dimensional Lorentzian flat
space with metric $\eta$ and
choose a system of $N$ coordinates $Y^A\,, \,\,\, A=0,\ldots,N-1,$
in the host-space ${\mathcal E}_{N}$.

To view the physical 4-D space-time as embedded in the host-space is
tantamount to saying that an arbitrary point $P$ in $\mathcal M$ can
be considered as a point of ${\mathcal E}_{N}$ as well. We thus
define an embedding by a set of $N$ functions $y^{A}(P)$, $A = 0,
..., N-1$, which represent the $Y$-coordinates of the space-time
point $P$. Note that these functions are scalars with respect to
coordinate changes on the space-time $\mathcal M$. Let us now choose
a system of four coordinates $x^\mu\,, \,\,\, \mu=0,1,2,3$ on the
physical space-time $\mathcal M$. The squared line element $ds^2$
between two infinitesimal points of $\mathcal M$ reads, with obvious
notations:
\begin{equation}
ds^2 = g_{\mu \nu} dx^{\mu} dx ^{\nu}; \label{eq:ds21}
\end{equation}
but the same squared line element can also be evaluated by viewing
both points as belonging to the host-space; this leads to
\begin{equation}
ds^2 = \eta_{A B} dy^{A} dy^{B}
\end{equation}
or
\begin{equation}
ds^2 = \eta_{A B} y^{A}_{,\mu} y^{B}_{,\nu} dx^{\mu} dx^{\nu},
\label{eq:ds22}
\end{equation}
where $y^{A}_{,\mu}$ denotes the partial differentiation of $y^{A}$
with respect to $x^{\mu}$. This partial derivative actually
coincides with the covariant derivative $y^{A}_{;\mu}$ of $y^{A}$
with respect to $x^{\mu}$ because, as noted earlier, the function
$y^{A}$ is a scalar respect to coordinate changes on $\mathcal M$.
Equating (\ref{eq:ds21}) and (\ref{eq:ds22}) delivers the important
relation:
\begin{eqnarray}\label{embed1}
\nonumber g_{\mu \nu} &=&\eta_{A B}
y_{\,\,\,;\mu}^A\,y_{\,\,\,;\nu}^B\\
&\equiv& \bm{y}_{;\mu}\cdot\bm{y}_{;\nu}\, ,
\end{eqnarray}
which is manifestly covariant with respect to coordinate changes on
$\mathcal M$.

\subsubsection{Existence of embeddings}

It is a well known result that a given Lorentzian (or Riemannian)
metric manifold can be embedded into a flat host space of higher
dimension. Constructive and existence theorems in the local
\cite{Car27,Fri65} as well as in the global sense
\cite{Nas56,{Cla70},{Gre70}} give conditions on the minimal
dimension of the host space, for closed and open manifolds (see also
\cite{And02}, and the references in the review \cite{Pav01}). The
minimal dimension of the host-space needed to embed locally a
generical 4-dimensional space-time is $N=10$. Usually less
dimensions are needed for vacuum space-times \cite{Fro59,Ros65}.

It has however been argued heuristically by Deser \emph{et
al.} \cite{Des76} that embeddings cannot \emph{a priori} be used with
profit by physicists. This conclusion essentially rests on an
intuition gained from studying the so-called trivial embedding of
$4$-D Minkovski space-time into itself, which cannot be deformed to
accomodate standard gravitational waves.
The way out of this possible problem is conceptually extremely
simple. It consists in working only with particular embeddings which
do admit deformations. This is where the notion of
freeness\cite{Nas56,{And02}} enters the picture.

\subsection{Free embeddings}
\label{subsec:free embeddings}

\subsubsection{Definitions}
\label{sssec:def}

Put simply, free, $q$-free and spatially free embeddings are three
particular classes of embeddings which share the common property of
being \emph{by definition} deformable to accommodate linear
variations of the metric tensor. Let us now present the
technicalities which motivate the three definitions we are about to
give.

Consider a given embedding of the form (\ref{embed1}) and let ${\bm
{\delta y}}$ be an arbitrary perturbation or deformation of this
embedding. We assume that the vectors ${\bm{y}_{; \mu}\,,\,\,\mu =
0, \ldots, 3}$ of the host space are linearly independent. Note that
this condition is necessary for the metric to be invertible, since
from eq.(\ref{embed1}), the linear dependence of these vectors would
imply directly the existence of a nonzero eigenvector of the metric
matrix $g_{\mu \nu}$ with zero eigenvalue.

Varying (\ref{embed1}), we obtain at first order in $\bm{ \delta
y}$:
\begin{equation}
\delta g_{\mu \nu } = 2 (\bm{y}_{; (\mu}\cdot \bm{\delta y})_{;
\nu)} - 2\, \bm{y}_{; \mu \nu} \cdot \bm{\delta y}\,.
\label{eq:deltag}
\end{equation}
The embedding variation is made up of two contributions, one tangent
to the 4-D space-time and the other one normal to it; we thus write
\begin{equation}
\bm{\delta y} = \delta W^\mu \bm{y}_{;\mu} \oplus \bm{\delta
y}_{\perp}\,.
\end{equation}
Equation (\ref{eq:deltag}) then becomes:
\begin{equation}\label{variationMetricSplit} \delta g_{\mu \nu } =
2 \,\delta W_{(\mu; \nu)} - 2\, \bm{y}_{; \mu \nu}\cdot \bm{\delta
y}_{\perp}\,.
\end{equation}
As (\ref{embed1}), (\ref{variationMetricSplit}) is manifestly
covariant with respect to coordinate changes on $\mathcal M$. The
vectors $\bm{y}_{;\mu \nu}\,,\,\,\mu,\nu=0,\ldots, 3 $ of the host
space define the \emph{second fundamental form} and are normal to
the 4-D embedded space-time: $\bm{y}_{;\mu \nu} \cdot
\bm{y}_{;\alpha} \equiv 0$ (see appendix \ref{appendix:covariant}).

Now, the embedding ${\bm y}$ will be useful physically if it can be
deformed to accommodate an arbitrary perturbation of the metric
${\delta g}$. This means that the physically useful embeddings are
those for which equation (\ref{variationMetricSplit}) can be solved
in ${\delta W_\mu}$ and $\bm{\delta y}_\perp$ for any arbitrarily
given $\delta g_{\mu \nu}$. Let us now introduce the definitions of
free, $q$-free and spatially $q$-free embeddings.

\noindent \textbf{Definition 1: Free embeddings.} An embedding
(\ref{embed1}) is said to be free if, for any metric perturbation
$\delta g_{\mu \nu}$ and any choice of tangential variation ${\delta
W_\mu}$, the 10 equations (\ref{variationMetricSplit}) can be solved
in the normal variations $\bm{\delta y}_{\perp}$.

This definition originated with Nash' work on embeddings of
Riemannian manifold and is now standard. Nash actually chose to work
with vanishing tangent variations and the normal variations can then
be obtained by algebraic methods only (see \cite{And02} for a modern
discussion). Locally, the dimension of the host space of Nash' free
embeddings must be $N\geq14.$

The material presented in the following sections (in particular, the
various action principles discussed in Section
\ref{sec:ActionPrinciples}) makes it natural to introduce two other
classes of embeddings, the $q$-free and spatially free embeddings.
These classes have never been considered by earlier authors and we
now give their definitions:

\noindent \textbf{Definition 2: $q$-free embeddings.} An embedding
(\ref{embed1}) is $q$-free if, for an arbitrary metric perturbation,
the 10 equations
(\ref{variationMetricSplit}) are equivalent
to:\\
(i) $q \geq 6$ linearly independent linear combinations of
(\ref{variationMetricSplit}) which can be solved in the normal
variations, no matter
what the tangential variations are, and\\
(ii) $10-q$ \, remaining equations that can be solved in the
tangential variations, independently of the solution obtained in
(i) for the normal variations. \\
\indent Notice that $10 \geq q \geq 6,$ because there are $4$
independent tangential variations and thus the number of equations
in (ii) must lie between $0$ and $4$. Note also that, by definition,
any free embedding is automatically a $10$-free embedding.

The host-space dimension for a $q$-free embedding must be $N \geq
q+4$ in order to accommodate the linearly independent
$4$-dimensional tangent space and the $q$-dimensional subspace of
the normal space which solve the equations in (i). Since $q\geq 6$,
we have $N \geq 10$.

\noindent \textbf{Definition 3: Spatially free embeddings.} An
embedding is spatially free if there exist a coordinate system
$(x^0, x^i)$ on the 4-D space-time in which the 6 vector fields
$\bm{y}_{; i j}\,,\,\,i,j = 1,2,3,$ are linearly independent.

Spatially free embeddings are particularly important when working in
the so-called $3 + 1$ formalism. They form a subclass of the
$q$-free embeddings. Indeed, let $\bm{y}$ be a spatially free
embedding and let $(x^0, x^i)$ be the coordinate system in which the
6 vector fields $\bm{y}_{; i j}\,,\,\,i,j = 1,2,3,$ are linearly
independent. In this coordinate system, the $(i,j)$ components of
(\ref{variationMetricSplit}) can be solved for $\bm{\delta
y}_\perp$. The remaining components of (\ref{variationMetricSplit})
are the $(0,\mu)$ components; taken together, they constitute a
system of four inhomogeneous first order differential equations for
the fields $\delta W_\mu,$ that can be solved by integration along
$x^0.$ Thus, any spatially free embedding is at least $6$-free,
hence $q$-free.

The three above definitions are relevant for physics if at least
some general relativistic space-times admit embeddings which are
either free, $q$-free or spatially free. We prove in the next section that this is so
by constructing explicit examples of free, $q$-free and
spatially free embeddings for several space-times of physical
interest, including a Schwarzschild black hole.
Whether all general relativistic space-times admit embeddings
in at least one of these three classes remains however an
open problem.

\subsection{Examples}\label{sec:Examples}
We now give explicit examples of embeddings of physically relevant
space-times belonging to the classes defined in the last section.

They represent thus the first examples in the literature of
embeddings whose deformations can, by construction, be properly
mapped to metric deformations, so that, for example, gravitational
waves can be described as embedding waves.

Let the space-time coordinates be denoted by $(x^0, x^j)$, where
$\{x^j, j=1,2,3\}$ are `spatial' coordinates.

We have developed a very simple method to construct $q$-free
embeddings for an interesting class of space-times, including the
flat and Schwarzschild $4$-dimensional space-times. The method
consists in splitting the host space as a direct sum of two flat
subspaces, (i) the `base' host space, with flat Euclidean metric
${\mathrm{diag}}(-1,\ldots,-1)$, and (ii) the `extra' host space,
with flat Lorentzian metric ${\mathrm{diag}}(1,-1,\ldots,-1)$ (one
and only one time-like direction). We will describe the contribution
of each subspace to the total embedding in two steps, followed by
the explicit examples.

\subsubsection{First Step: The base embedding}
\label{sec:firstStep} On the base host space, we define a `base
embedding' $\bm{Z}(x^j)$, depending only on the spatial coordinates,
with the following properties:

\begin{enumerate}
    \item[P1.] The metric induced by $\bm{Z}$ is flat: $\bm{Z}_{,j}\cdot\bm{Z}_{,k} = -\delta_{j
    k}$
    \item[P2.] The $9$ vectors $\{\bm{Z}_{,j}, \bm{Z}_{,k l}\}$ are
    linearly independent for all $x^j \in \mathcal{B}$,
where $\mathcal{B}$ is a given $3$-dimensional subset of
$\mathbb{R}^3$.
\end{enumerate}
 Among the several types of base embeddings that can be
constructed we have chosen the following $11$-dimensional one, for
its simplicity:
\begin{equation}\label{embedding-example-flat-z}
\bm{Z}(x^j) = \left(\begin{array}{l}
{f_1(\xi_1\,x^1)}\\
{f_2(\xi_2\,x^2)}\\
{f_3(\xi_3\,x^3)} \\
\cos(g_1(\xi_1\,x^1))\,
   \cos (g_2(\xi_2\,x^2))\,
   \cos (g_3(\xi_3\,x^3))\\
\cos (g_1(\xi_1\,x^1))\,
   \cos (g_2(\xi_2\,x^2))\,
   \sin (g_3(\xi_3\,x^3))\\
\cos (g_1(\xi_1\,x^1))\,
   \sin (g_2(\xi_2\,x^2))\,
   \cos (g_3(\xi_3\,x^3))\\
\cos (g_1(\xi_1\,x^1))\,
   \sin (g_2(\xi_2\,x^2))\,
   \sin (g_3(\xi_3\,x^3))\\

   \sin (g_1(\xi_1\,x^1))\,\cos (g_2(\xi_2\,x^2))\,
   \cos (g_3(\xi_3\,x^3))\\

   \sin (g_1(\xi_1\,x^1))\,\cos (g_2(\xi_2\,x^2))\,
   \sin (g_3(\xi_3\,x^3))\\

   \sin (g_1(\xi_1\,x^1))\,
   \sin (g_2(\xi_2\,x^2))\,\cos (g_3(\xi_3\,x^3))\\
\sin (g_1(\xi_1\,x^1))\,
   \sin (g_2(\xi_2\,x^2))\,
   \sin (g_3(\xi_3\,x^3))\,,
\end{array}\right)
\end{equation}
where the functions $f_j, g_j$ and the parameters $\xi_j$ are chosen
so that $\bm{Z}$ satisfies properties P1 and P2.
Property P1 is ensured if there exist real functions $u_j(s)$ such that:
\begin{equation}\label{eq:expon}
f_j'(s)+ i\,g_j'(s) = \xi_j^{-1} \,{\mathrm{e}}^{i\, u_j(s)}, \quad
j=1,2,3,
\end{equation}
where $i^2= -1$ and prime denotes derivative with respect to $s$.
Property P2 is ensured if the real functions $u_j(s)$ satisfy:
\begin{eqnarray}\nonumber
&0 < u_j(s)<\pi,\quad j=1,2,3 \\
&\Delta \equiv {\left({v_1}{v_2}{v_3}\right)}^2 +
{\left({w_1}{v_2}{v_3}\right)}^2 +{\left({v_1}{w_2}{v_3}\right)}^2+
{\left({v_1}{v_2}{w_3}\right)}^2>0\nonumber
\end{eqnarray}
where $v_j \equiv \xi_j u_j'(\xi_j x^j)$ and $w_j\equiv\sin^2
u_j(\xi_j x^j)$, for $j$ fixed.
A particular solution of the previous inequalities is
$$u_1(s) =
\frac{\pi}{4}, \quad u_2(s) = u_3(s)  = \frac{1}{4}\left(\pi +
\arctan s\right).$$

With this solution, $\bm{Z}$ is a base embedding on
$\mathcal{B} = \mathbb{R}^3 \backslash \{x^1\in \mathbb{R},x^2=\pm
\infty,x^3=\pm \infty\}$ and this embedding is not self intersecting (\emph{i.e.} for any two sets of 3 spatial coordinates $p$ and $q$,
$\bm{Z}(p) = \bm{Z}(q) \iff p = q\,).$

\subsubsection{Second step: the extra embedding functions}
It is easy to show that, by adding extra embedding functions to this
base embedding, the general embedding
$\bm{y}(x^0,x^j) \equiv \bm{Y}(x^0,x^j) \oplus \bm{Z}(x^j)$
satisfies automatically the following properties:
\begin{enumerate}
    \item[P1'.] The metric induced by $\bm{y}$ decomposes as:
\begin{equation}\label{eq:splitMetric}
ds^2 = \bm{Y}_{,\mu}\cdot\bm{Y}_{,\nu}dx^\mu dx^\nu -\delta_{j
k}dx^j dx^k
\end{equation}
    \item[P2'.] The embedding $\bm{y}(x^\mu)$ is spatially free for $x^j \in \mathcal{B}$.
\end{enumerate}

Appropriate choices of these extra embedding functions $\bm{Y}$ generate
spatially free embeddings of the flat Minkovski space-time and
of the Schwarzschild black hole.

\subsubsection{First example:
Flat 4-D space-time embedded into flat (1+11)-D host space}
\label{FirstExample}

Here, to accommodate the extra embedding, only one extra dimension
(necessarily time-like) is needed. The extra embedding function is
$\bm{Y}_{\mathrm{Flat}}(x^\mu) = \left( x^0\right)$. The
($(1+11)$-dimensional) host space metric is $\eta_{A B} =
{\mathrm{diag}}(1,-1,\ldots,-1)$, and it follows from eq.(\ref{eq:splitMetric})
that the induced space-time metric is $g_{\mu \nu} =
{\mathrm{diag}}(1,-1,-1,-1).$ This $12$-dimensional
embedding is $6$-free for $x^j \in \mathcal{B} = \mathbb{R}^3
\backslash \{x^1\in \mathbb{R},x^2=\pm \infty,x^3=\pm \infty\}.$

\subsubsection{Second example: Schwarzschild 4-D black hole (Kerr-Schild coordinates)
into (1+14)-D host space}

\begin{figure}[h]
\begin{center}
\epsfxsize=110pt
 \epsfbox{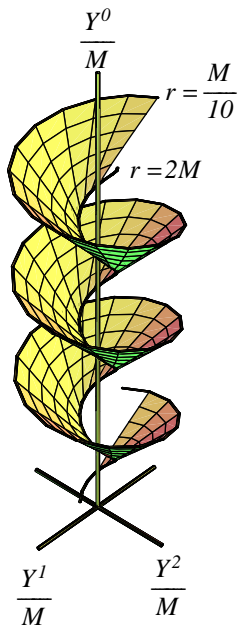}
\end{center}
\caption{\label{fig:BH1} Schwarzschild
black hole embedding, eq.(\ref{plot}). The
helicoidal surface shown represents a piece
of the interior zone $r<2M$, bounded
by the wider, space-like helix $r=0.1 M$ and
by the light-like helix $r=2M$ (the event horizon).}
\end{figure}

\begin{figure}[h]
\begin{center}
\epsfxsize=60pt \epsfbox{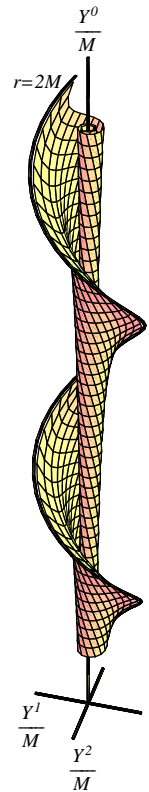}
\end{center}
\caption{\label{fig:BH2} Schwarzschild
black hole embedding, eq.(\ref{plot}):
exterior zone $r\geq 2M$. The spiral
surface folds and folds indefinitely as $r\to
\infty$,
approaching the axis $Y^{0}/M$, never reaching it.
The upgoing helix bounding the surface corresponds to the event
horizon $r=2M$.}
\end{figure}

In this case, to accommodate the extra embedding we need $3$ extra
dimensions, one time-like and two space-like.
The metric components $g_{\mu\nu}(x^0,x^1,x^2,x^3)$ of this black
hole in Kerr-Schild coordinates do not depend explicitly on $x^0$,
and read \cite{Debb}:
$$g_{00} = 1 - \frac{2\,M}{r},\,\, \quad g_{0 j} =
-\frac{2\,M\, {x^j}}{r^2},\,\, \quad g_{j k} = -\delta_{j k}
-\frac{2\,M\,{x^j \,x^k}}{r^3}\,,$$ where $j,k = 1,2,3,$ and
 $r=\sqrt{(x^1)^2+(x^2)^2+(x^3)^2}.$
By looking at eq.(\ref{eq:splitMetric}), this black-hole metric is
obtained if we use the following extra embedding:
\begin{equation}\label{plot}
\begin{array}{lcl}
Y_M^{0}(x^\mu) & = & x^0 - M h_1(\frac{r}{M})\\
{Y}_M^{1}(x^\mu)&=&\frac{6 \sqrt{3} M}{\zeta} \sqrt{\frac{2 M}{r}}
\sin \left(\frac{\zeta}{6 \sqrt{3} M}\left(x^0 - M
h_2(\frac{r}{M})\right)\right)\\
Y_M^{2}(x^\mu)&=&\frac{6 \sqrt{3} M}{\zeta} \sqrt{\frac{2 M}{r}}
\cos \left(\frac{\zeta}{6 \sqrt{3} M}\left(x^0 - M
h_2(\frac{r}{M})\right)\right)\,.
\end{array}
\end{equation}
Each spatial coordinate $x^j$ ranges from $-\infty$ to $\infty.$ The
parameter $\zeta$ must satisfy $\zeta^2 \geq 1.$ Finally, the
functions $h_1(s)$ and $h_2(s)$ must solve the differential equations: $
h_1'(s)= \frac{2(1+h_2'(s))}{s},\quad h_2'(s)= \frac{2\,s -
{\sqrt{s^4 + {(54 - 27\,s)\,s\,{{\zeta }^{-2}}}}}}
  {\left( s -2 \right) \,s}.$
They are well-behaved for all $s>0$, even near the horizon ($s=2$).
Analytical expressions for the functions $h_1(s)$ and $h_2(s)$
can be obtained
when $\zeta = 1:$ Figs.\ \ref{fig:BH1} and \ref{fig:BH2} have been
constructed using these expressions.

The extra embedding functions $Y_M^{0},
Y_M^{1}, Y_M^{2}$ account for the causal
structure of the black hole. After rescaling the time $x^0$, the radial coordinate $r$
and the embedding functions in units of $M$, we plot
them from eqs.(\ref{plot}) in Figs.\ \ref{fig:BH1} and \ref{fig:BH2}
(interior and exterior region, respectively)
as two-dimensional surfaces parameterized by $(\tau=x^0/M,
s=r/M).$
These are helicoidal surfaces with a pitch $\Delta \tau = \pi \sqrt{54}$.
We remark that the interior and exterior regions are smoothly connected.
The black hole metric is equal to the
 metric induced by the plotted
surface plus the (flat) metric
$\bm{Z}_{,j}\cdot\bm{Z}_{,k}= - \delta_{j k}$
induced by the base embedding (not plotted).
The event horizon $r=2 M$, denoted in both figures, is an
upgoing helix subtending an angle of
$\pi/4$ with respect to the rescaled axis
$Y^{0}/M$.

The interior region $0<r<2 M$ is
represented in Fig.\ \ref{fig:BH1}.
We can see the event horizon $r=2 M$ as
a light-like helix which is close to the vertical
axis. The space-like
helix $r=0.1 M$ is bounding the figure. The
actual surface extends to infinity as $r\to
0,$ subtending asymptotically an angle of
$\pi/4$ with respect to the rescaled axis $Y^{0}/M$ as
$r \approx 0:$ all physical trajectories
get trapped in the interior region,
approaching the `helix at infinity' $r=0$. This
graphical interpretation is possible since
the plotted surface gives the main contribution to the black hole
metric in the region $r \approx 0$. Indeed, the
(flat metric) contribution coming from the base embedding
(not plotted) is proportional to $r^2/M^2$ and is therefore
negligible.

The exterior region $r > 2 M$ is
represented in Fig.\
\ref{fig:BH2} by the spiral surface that approaches the vertical axis as
$r\to \infty$, folding and
folding indefinitely, never reaching the
axis. Notice that in this region, the above mentioned base embedding's contribution
(not plotted) to the black hole metric, being proportional to $r^2/M^2$,
 is much more important than the \emph{spatial}
contribution coming from the plotted surface, so that this exterior region plot
 gives only partial information about the causal structure of the black hole.

\subsubsection{Third example: Waves around the embedding of flat space-time}
\label{sec:example-waves} Let us check that
the $6$-free embedding of flat space-time
introduced in section \ref{FirstExample} allows for the representation of
gravitational waves as embedding waves  by computing explicitly the
perturbation of this embedding that corresponds to
the well-known \cite{W84a} plane wave:
$\delta g_{\mu \nu} = L_{\mu \nu}
{\mathrm{e}}^{i p_\alpha x^\alpha},$ with $p_\mu p^\mu = 0$, $L_{0
\mu}= 0$ and $L_{j k} p^k = 0.$
We solve
eq.(\ref{variationMetricSplit}) for the embedding variations by
first taking vanishing tangent variations $\delta W_\mu = 0$.
We then split the
normal variations as the embedding itself:
$\bm{\delta y}_\perp = \bm{0} \oplus \bm{\delta
Z}_\perp$, where $\bm{\delta Z}_\perp \equiv \bm{Z}_{,j k} \delta
f^{j k}$; $\delta f^{1 2}$ and $\delta f^{1 1}$ are explicitly given by:
\begin{equation}\label{eq:solsixeqs}
\begin{array}{rcl}
\delta f^{1 2} & = & -\frac{{\delta
g}_{1 2}}{4 w_1 w_2}, \\
{{(2\, \Delta)}}\,\delta f^{1 1}&=&-{
  {{\delta g}_{1 1}}\left(\left({{v_2}{v_3}}\right)^2
  +
     \left({{v_2}{w_3}}\right)^2 + \left({{w_2}{v_3}}\right)^2 \right)}\\
&+&{{\delta g}_{2 2}\,w_1{w_2}{{v_3}}^2 +
  {{\delta g}_{3 3}}\,w_1{{v_2}}^2{w_3}
  },
\end{array}
\end{equation}
and the other components are obtained by cyclic permutations of
$(1,2,3)$. The functions $v_i, w_j$ and $\Delta$ were defined in Section
\ref{sec:firstStep}.

The embedding waves are thus simply plane waves, modulated with some
smooth functions related to this particular embedding.

\section{Action Principles for embedding theory}
\label{sec:ActionPrinciples}

We have introduced in the previous sections the (purely kinematical)
concept of free embedding, which is \emph{by definition} deformable
to accommodate linear variations of the metric tensor, and thus
gravitational waves. Our next goal, therefore, is to find a
\textbf{satisfactory action principle} for general relativity in
terms of embedding variables, in order to obtain what we will call
\textbf{`free embedding field theory for gravity'}. This theory that
requires host space dimension $N\geq 14$, turns out to be equivalent
to GR, not only at the levels of EOM but also for perturbations
(waves around the general solution). We then consider the
\textbf{`$q$-free embedding theory for gravity'}, that allows us to
reduce the minimal dimension of the host space from $14$ to $10$.

Finally, we introduce a third, alternative family of action
principles that depend on Lagrange multipliers. The metric
variations and embedding variations are independent and their
relation (eq.(\ref{embed1})) is given as an EOM. These alternative
theories are also  shown to be equivalent to usual GR, both at the
levels of EOM and of perturbations (waves around the general
solution), provided that the embedding is $q$-free.

\subsection{Hilbert action in terms of
embedding variables}\label{subsec:motivation} Let us denote the
$4$-D space-time manifold and its $3$-D boundary by ${\mathcal{M}}$
and $\partial {\mathcal{M}}$ respectively and let $N$ be the
dimension of the host space. Consider the usual \cite{W84a} Einstein
action principle but written in terms of embedding variables
\begin{equation}\label{actionRT} S[\bm{y}] = \frac{1}{8\pi
G}\int_{\mathcal{M}} R[g[\bm{y}]]\,\sqrt{-|g[\bm{y}]|}\,d^4 x +
\frac{1}{4\pi G}\int_{\partial \mathcal{M}}
K[\bm{y}]
\end{equation}
where $g[\bm{y}]$ means that $g$ is replaced everywhere by its
definition, eq.(\ref{embed1}). The scalar curvature $R[g[\bm{y}]]$
is given explicitly by $R = \bm{y}_{;\alpha}^{\!\!\quad;\alpha}
\cdot \bm{y}_{;\beta}^{\!\!\quad;\beta} - \bm{y}_{;\alpha \beta}
\cdot \bm{y}^{;\alpha \beta}.$ Note that the scalar curvature is
quadratic and of second order only in the embedding functions
$\bm{y}$ (see appendix \ref{appendix:covariant}). Finally,
$K[\bm{y}]$ is the trace of the extrinsic curvature of the boundary.
The $K$-boundary term is here, as usual, \cite{W84a} to cancel
boundary terms that depend on the normal derivatives of $g_{\mu
\nu}$, and allows to impose the boundary conditions $\delta g_{\mu
\nu} =0$.

We thus assume from now on the vanishing boundary conditions
(v.b.c.) $\delta g_{\mu \nu} = 0$ on $\partial{\mathcal{M}}$. These
boundary conditions explicitly read, using
eq.(\ref{variationMetricSplit}),
\begin{equation}\label{vbc} 0=
2 \,\delta W_{(\mu; \nu)} - 2\, \bm{y}_{; \mu \nu}\cdot \bm{\delta
y}_{\perp}\,.
\end{equation}
Let us consider the variation of action (\ref{actionRT}), in two
steps. First, we obtain in the standard way \cite{W84a}
\begin{equation}\label{eq:delta action 1}
\delta S = -\frac{1}{8\pi G}\int_{\mathcal{M}} G^{\mu \nu} \delta
g_{\mu \nu} \sqrt{-|g|} d^4 x\,,
\end{equation}
where $G^{\mu \nu}[\bm{y}]$ is the Einstein tensor computed in terms
of the embedding functions (see appendix \ref{appendix:covariant})
and $\delta g_{\mu \nu}$ is the variation of the metric given in
terms of the variation of the embedding functions in
eq.(\ref{variationMetricSplit}). Next we rewrite this variation of
the action explicitly in terms of the variations of the embedding
functions. Using the twice contracted Bianchi identity
$G_{\,\,\,\,;\nu}^{\mu \nu} \equiv 0\,,$ we get a boundary plus a
bulk term:
\begin{widetext}
\begin{equation}\label{Var-action-y}
\delta S = -\frac{1}{4\pi G}\int_{\partial\mathcal{M}} \delta W_\mu
G^{\mu \nu} d \Sigma_\nu + \frac{1}{4\pi G}\int_{\mathcal{M}}
G^{\mu \nu} \,\bm{y}_{; \mu \nu}\,\cdot\bm{\delta y}_\perp
\sqrt{-|g|} d^4 x \\,
\end{equation}
\end{widetext}
where $d \Sigma_\nu$ is the normal unitary 3-volume element on the
boundary. The above boundary term comes from the differential
dependence of the metric in terms of the embedding functions. Note
that it has nothing to do with the $K[\bm{y}]$-boundary term in
eq.(\ref{actionRT}).

\subsection{Free embedding theory equivalent to
General Relativity} \label{sec:freeEmbTheory} We assume in this
section the simplest case which is that the embedding is free
throughout the manifold and at its boundary. From local existence
theorems, this will require the dimension of the host space to be
$N\geq 14$. By definition of free embedding (see section
\ref{sssec:def}), the $10$ $N$-dimensional host-space vectors
$\{\bm{y}_{; \mu \nu}, \quad \mu, \nu = 0, \ldots, 3\}$ are linearly
independent.

In this simple case, we can consistently take $\delta W_\mu = 0$
throughout the manifold. The boundary conditions on the embedding
functions (\ref{vbc}) now imply, because of the freeness of the
embedding, $\bm{\delta y}_\perp = 0$ on the boundary.

Using eq.(\ref{Var-action-y}), under arbitrary normal embedding
variations $\bm{\delta y}_\perp$ in the bulk we obtain the following
Euler-Lagrange equations:
\begin{equation}\label{eomRT}
\qquad G^{\mu \nu}[\bm{y}] \,\bm{y}_{; \mu \nu}=0\,
\end{equation}
which, taking into account embedding freeness in the bulk, imply
$G^{\mu \nu}[\bm{y}] = 0,$ which are the standard Einstein equations
but written in terms of the embedding functions.

It is worthwhile at this point to warn the reader against a common
misinterpretation. In connection with string-like models with
arbitrary $N$, the equations for the minimal hypervolume
membrane,\footnote[1]{The action for the minimal hypervolume
membrane is $S_h[y^A]=\int_{\mathcal{M}}\sqrt{-|g[y]|}d^4 x$.} $
g^{\mu \nu} \,\bm{y}_{; \mu \nu}=0\,,$ are correctly understood as
constraints on the second fundamental form $\bm{y}_{; \mu \nu}$
\cite{Reg75,{Mai89}}. But the analogy of these equations with
eqs.(\ref{eomRT}), with $G^{\mu \nu}$ replacing $g^{\mu \nu}$ leads
to a widespread error: to also understand
 eqs.(\ref{eomRT}) as constraints
on the second fundamental form. For free embeddings this is clearly
a wrong interpretation since, as we have shown, when a free
embedding solves eq.(\ref{eomRT}), it follows $G^{\mu \nu} = 0$ in a
perfectly consistent way. On the other hand, for the minimal
hypervolume membrane this interpretation is correct, since if a free
embedding solved the corresponding equation, then the contradictory
result $g^{\mu \nu}=0$ would follow \footnote[2]{It remains an open
problem if there exist $q$-free embeddings that solve the membrane
equations: a careful study must be made on the membrane action
principle and the boundary conditions on the embedding variations.}.

\subsection{$q$-free embedding theories equivalent to General Relativity}
\label{sec:q-freeEmbTheory} The earliest work we know on the
variational principle for gravitation in terms of embedding
variables is the `gravitation $\grave{a}\,\, la$  string' due to
Regge and Teitelboim \cite{Reg75} (RT). This approach is well-known
\emph{not} to be equivalent to general relativity
\cite{Reg75,{Des76},{Mai89}}. The reason for this is as follows. By
requiring isometric embedding only, the authors set $N=10$ as the
dimension of the flat host space. They use essentially the action
(\ref{actionRT}) as a starting point, and the equations of motion
are again (\ref{eomRT}). However, in this case the equations no
longer imply the Einstein equations since the embedding cannot be
free, for the dimension of the host space is $N=10 < 14.$ Indeed,
there are only $6$ independent EOM.

The free embedding theories introduced in the previous section are
equivalent to GR but require a host space of dimension $N\geq 14$ .
The following question therefore naturally arises: `are there
embedding theories equivalent to GR, based on a host space of
dimension $10 \leq N <14$?'

\subsubsection{$6$-free $N=10$ case}
\label{sssec:6freeN=10} In this section we consider the most
challenging case, $N=10$. In other words, is it possible to modify
$N=10$ RT theory and make it equivalent to GR? We now show that the
only way to make a $N=10$ RT theory equivalent to GR is to require
the embedding to be $6$-free together with appropriate boundary
conditions on the embedding variations.

In order to be equivalent to GR, a theory should have the same
number (ten) of independent EOM. However, as pointed out by Regge
and Teitelboim \cite{Reg75}, Deser \emph{et al.} \cite{{Des76}},
Pav\v{s}i\v{c} \cite{Pav85}, and Franke and Tapia \cite{Fra92}, it
follows from $\bm{y}_{; \mu \nu}\cdot \bm{y}_{;\lambda} = 0$ that
the second fundamental form $\bm{y}_{; \mu \nu}$ has at most $6$
independent components (instead of ten). There are therefore at most
$6$ independent EOM (\ref{eomRT}). Note that they are \emph{exactly}
$6$ independent equations (by definition 2, section \ref{sssec:def})
when the embedding is $6$-free. What happened to the other 4?

The answer comes from the fact that the RT boundary conditions do
not allow  for an arbitrary variation $\delta g_{\mu \nu}$ in the
bulk, which is not consistent with the standard derivation of the
Einstein equations from the action principle (\ref{actionRT}) (see
eq.(\ref{eq:delta action 1})). The possibility of having arbitrary
$\delta g_{\mu \nu}$ in the bulk depends on the boundary conditions
imposed on the embedding functions in the theory. But, as we will
see below, the naive RT choice of v.b.c.\ on the \emph{variations of
embedding functions} and their normal derivatives imposes
constraints on the bulk metric variations, independently of the
$6$-freeness property. However, an arbitrary variation $\delta
g_{\mu\nu}$ in the bulk is allowed when the boundary conditions: (a)
$\delta W_\mu$ arbitrary and (b) $\bm{\delta y}_\perp$ given by
eqs.(\ref{vbc}) on the boundary, are used instead of the RT v.b.c.

To give a specific example, consider variations around a $N=10$,
$6$-free and spatially free embedding of an open neighborhood within
the flat space-time $\mathbb{R}\times \mathbb{R}^3$. Here the
relevant boundaries for the action principle are the space-like
boundaries $x^0=0,\,x^0= 1$. We assume for simplicity the following
property for the embedding functions: $\bm{y}_{; 0 \mu} =0 $
throughout the open neighborhood. Now suppose that both $\bm{\delta
y}_{\perp}$ and $\delta W_{\mu}$ are zero on the boundaries.

Following the general equations (\ref{variationMetricSplit}) there
are, in the bulk, 4 differential equations we must solve for the
tangent embedding variations: $\delta g_{0 \nu} = 2 \delta W_{(0,
\nu)}\,,\quad \nu=0,1,2,3, $ and 6 algebraic equations for the
normal embedding variations: $\delta g_{i j} = 2 \delta W_{(i, j)} -
2 \bm{y}_{,i j} \cdot\bm{\delta y}_{\perp} \,,\quad i,j =1,2,3. $
The last 6 equations imply $\delta g_{i j}=0$ on the boundaries,
with no condition in the bulk. However, the first 4 equations give
constraints: the $\nu=0$ component implies $\delta
W_0(x^0,x^i)|_{x^0=0}^1 = \frac{1}{2}\int_0^{1}\,\delta g_{0
0}(z,x^i)\,dz,$ and the $\nu =j$ components imply
 $\delta
W_j(x^0,x^i)|_{x^0=0}^1 = \int_0^{1}\,\delta g_{0 j}(z,x^i)\,dz\, -
\frac{\partial}{\partial x^j}\int_0^{1}\,dz\,\int_0^z\,d s\,\delta
g_{0 0}(s,x^i)\,.$ But these expressions are zero for the v.b.c.\ on
the embedding variations, so we obtain $\int_0^{1}\,\delta g_{0
0}(z,x^i)\,dz\, = 0,$ $\int_0^{1}\,\delta g_{0 j}(z,x^i)\,dz\, =
\frac{\partial}{\partial x^j}\int_0^{1}\,dz\,\int_0^z\,d s\,\delta
g_{0 0}(s,x^i).$\\

Conversely (see eqs. (\ref{variationMetricSplit})), the assumption
of arbitrary metric variations in the bulk imply that, at space-like
portions of the boundary, all the $4$ embedding tangent variations
$\delta W^\mu$ must be arbitrary. Recalling that the action
variation (\ref{Var-action-y}) depends on these tangent variations,
we are led to $4$ extra Euler-Lagrange equations on the boundary,
that will turn out to be the usual constraints of GR, see
eq.(\ref{RTEulerLagrBoundary}) below.

We now turn to the proof that, in the more general case of curved
space-times embedded in a spatially free manner, these extra
equations will be enough to show the equivalence of the Einstein
equations with the Euler-Lagrange equations obtained from the action
(\ref{actionRT}) under the new boundary conditions: $\delta W_\mu$
arbitrary and $\bm{\delta y}_\perp$ given by eqs.(\ref{vbc}), which
admit solutions because of spatial freeness. For simplicity, we will
assume the following properties for the space-time and the
embedding: (i) the space-time is globally hyperbolic, which allows
to define a global time coordinate, and (ii) the embedding is
$6$-free and, with respect to the latter time coordinate, it is
spatially free (see Definition 3 in section \ref{sssec:def}). We
define the resulting theory as a $6$-free embedding theory of
gravity.

Using the general equation (\ref{Var-action-y}) for the variation of
the action, we consider arbitrary variations $\delta W^\mu$ on the
boundary, and arbitrary variations $\bm{\delta y}_\perp$ in the
bulk. The Euler-Lagrange equations for this theory are thus:
\begin{equation}\label{RTEulerLagrBoundary}
{\mathrm{On}\,\,
\mathrm{{space}}\mathrm{-}\mathrm{{like}}\,\,\mathrm{portions \,\,
of\,\,{\partial\mathcal{M}}:}} \quad n_\mu G^{\mu \nu}=0,
\end{equation}
where $n_\mu$ is the unit normal to the boundary, and
\begin{equation}\label{RTEulerLagrBulk}
   {\mathrm{On \,\,{\mathcal{M}}:}} \quad G^{\mu \nu} \bm{y}_{; \mu
   \nu}=0\,.
\end{equation}
Using the appropriate time slicing, which exists in globally
hyperbolic space-times, the Euler-Lagrange equations at the
space-like boundary are $G^{0 \mu}|_{t = t_0} = 0$. On the other
hand, spatial freeness of the $10$-dimensional embedding implies the
relations $\bm{y}_{; 0 \mu} = A_\mu^{i j}\bm{y}_{; i j}$, where
$A_\mu^{i j}$ are tensor fields defined by the embedding. Therefore
the bulk eqs.(\ref{RTEulerLagrBulk}) become
\begin{equation}\label{RTEulerLagrBulk-1}
G^{i j} + G^{00} A_0^{i j} + 2\, G^{0 k} A_k^{i j} = 0\,.
\end{equation}
Using this equation we can rewrite $G_{\,\,\,\,;\nu}^{\mu \nu}
\equiv 0$ as a system of four first order, homogeneous partial
differential equations for the unknowns $G^{0 \mu}.$ Recalling that
the initial condition is $G^{0\mu}|_{t = t_0} = 0,$ we obtain $G^{0
\mu} =0$ in the bulk, and then eqs.(\ref{RTEulerLagrBulk-1}) imply
$G^{\mu \nu} =0$ in the bulk. We have thus shown the equivalence
between GR and $6$-free, spatially free embedding theory.

\subsubsection{$6$-free $N>10$ case}
\label{sssec:6freeN>10} The case $N>10$ corresponds to the addition
of extra embedding functions to the theory, and all derivations
above apply with minimal obvious modifications.

\subsubsection{Waves in $6$-free embedding
theory} We now explicitly demonstrate that the above embedding
theory linearized about the solution of the EOM $G^{\mu \nu} = 0$ is
equivalent to standard linearized GR, and thus contains the standard
gravitational waves. To wit, we compute the variations of
eqs.(\ref{RTEulerLagrBoundary}--\ref{RTEulerLagrBulk}), and evaluate
them on the general solution, which satisfies the EOM $G^{\mu \nu}
=0$. In the rest of this subsection the covariant derivative
corresponds to the metric satisfying the EOM. We obtain:
\begin{eqnarray}\label{deltaRTEulerLagrBoundary}
   {\mathrm{On \,\,\,\,{\partial\mathcal{M}}:\,\,\,}}& n_\mu \delta G^{\mu
   \nu}&=0\,,\\
\label{deltaRTEulerLagrBulk}
   {\mathrm{On \,\,\,\,{\mathcal{M}}:\,\,\,}}& \delta G^{\mu \nu} \bm{y}_{; \mu
   \nu}&=0\,.
\end{eqnarray}
The tensor $\delta G_{\mu \nu}$ is covariantly conserved when the
EOM hold. This can be deduced from the variation of the identity
$G^{\mu \nu}_{\,\,\,\,; \nu}\equiv 0$. We get $(\delta G^{\mu
\nu})_{; \nu} + (\delta \Gamma_{\,\,\,\nu \lambda}^\mu) G^{\nu
\lambda}+(\delta \Gamma_{\,\,\,\nu \lambda}^\nu) G^{\mu \lambda} =
0$ and, using the EOM $G^{\mu \nu}=0,$  $(\delta G^{\mu \nu})_{;
\nu} =0$ follows. Then, the tensor $\delta G$ satisfies the same
equations as the tensor $G$. Repeating the arguments already used
above we conclude that for spatially free $6$-free embeddings the
solution is $\delta G^{\mu \nu}=0$, just as expected for the
standard perturbation theory of the Einstein equations in GR.
However, this time one has to compute the variations with respect to
the embedding variables. Though it is possible to write down the
resulting equations, in our opinion it gives no further insight to
do it. Instead we notice that the perturbations of the embedding
functions \emph{must} propagate in a proper way, because of the
proved correspondence that exists, for spatially free $6$-free
embeddings, between the variations of the metric and those of the
embedding, eqs.(\ref{variationMetricSplit}). We refer the reader to
the explicit example of waves around flat space-time in section
\ref{sec:example-waves}.

\subsubsection{The case of $q$-free embeddings with  $6<
q < 10$}

This case is intermediate between the $N\geq 14$ free (i.e.,
$10$-free) case (section \ref{sec:freeEmbTheory}) and the $N \geq
10$ $6$-free case (sections \ref{sssec:6freeN=10},
\ref{sssec:6freeN>10}).

In this case the EOM (\ref{eomRT}) are exactly $q>6$ equations (by
definition 2, section \ref{sssec:def}). Therefore there will be only
$10-q<4$ missing equations. These equations will be obtained as
boundary equations like eqs.(\ref{RTEulerLagrBoundary}), by assuming
arbitrary boundary conditions on $10-q$ out of the four $\delta
W_{\mu}$. Following the same type of arguments as those between
eqs.(\ref{RTEulerLagrBoundary}) and eqs.(\ref{RTEulerLagrBulk-1}),
we can conclude that if the embedding is spatially free and $q$-free
with $N\geq 4+q$, the EOM are equivalent to GR.

\subsection{Action Principles with independent metric and embedding functions}

This section deals with a class of action principles where the
metric $g_{\mu \nu}$ and the embedding functions $\bm{ y}$ are
considered independent.  Relation (\ref{embed1}) between metric and
embedding therefore appears as an EOM.

Consider the following action
\begin{widetext}
\begin{eqnarray}\nonumber
S_n[g,y,\lambda] &=& \frac{1}{8\pi G}\int_{\mathcal{M}} d^4 x\,\sqrt{-|g|}R[g] + \frac{1}{4\pi G}\int_{\partial \mathcal{M}} K[g]\\
\label{actionN}&+& \frac{1}{8\pi G}\int_{\mathcal{M}} d^4
x\,\sqrt{-|g|}\lambda^{\mu_1 \nu_1 \ldots \mu_n \nu_n}\left(g_{\mu_1
\nu_1}-\bm{y}_{,\mu_1}\cdot\bm{y}_{,\nu_1}\right)\ldots\left(g_{\mu_n
\nu_n}-\bm{y}_{,\mu_n}\cdot\bm{y}_{,\nu_n}\right),
\end{eqnarray}
\end{widetext}
where $R[g], K[g]$ are the Ricci scalar and the extrinsic curvature
of the boundary ${\partial \mathcal{M}}$, both  expressed in terms
of the metric,  and $\lambda$ is a Lagrange multiplier.

\subsubsection{The case $n=1$}
This case has already been presented in the literature \cite{Des76},
with v.b.c.\ on the embedding variables; these b.c. make the theory
inequivalent to GR. Let us consider now the action (\ref{actionN})
in the context of free and $q$-free embeddings and allow for
arbitrary tangent variations $\delta W_{\mu}$ of the embedding.

The variation of the $S_1$ reads:
\begin{widetext}
\begin{eqnarray}\nonumber
({8\pi G})\, \delta S_1 &=& \int_{\mathcal{M}} \left[\delta
\lambda^{\mu \nu} \left(g_{\mu \nu} -
\bm{y}_{,\mu}\cdot\bm{y}_{,\nu}\right) + \delta g_{\mu \nu}
\left(\lambda^{\mu \nu} + \frac{1}{2}g^{\mu \nu}\lambda^{\alpha
\beta}\left(g_{\alpha \beta}-
\bm{y}_{,\alpha}\cdot\bm{y}_{,\beta}\right)- G^{\mu \nu}[g]\right)
  \right]\sqrt{-|g|}
d^4 x
\\
\label{Var-action-1-degree} &-&2\int_{\partial\mathcal{M}} \delta
W_\mu
 \lambda^{\mu \nu} d
\Sigma_\nu+2\int_{\mathcal{M}} \left[\delta W_\mu \lambda_{\quad
;\nu}^{\mu \nu} + \bm{\delta y}_\perp \cdot \bm{y}_{; \mu
\nu}\lambda^{\mu \nu}\right]\sqrt{-|g|} d^4 x \,,
\end{eqnarray}
\end{widetext}
where the Einstein tensor $G^{\mu \nu}[g]$ is viewed as a
functionnal of the metric. Remember that $g_{\mu \mu}$ and ${\bm y}$
are treated as independent at this stage; their variations are
therefore independent too and eq. (\ref{Var-action-1-degree}) thus
delivers three Euler-Lagrange equations in the bulk. The
Euler-Lagrange equation with respect to the Lagrange multiplier
$\lambda^{\mu \nu}$ gives, as a constraint, the relation between the
metric and the embedding functions, eq.(\ref{embed1}); replacing
this equation into the Euler-Lagrange equation with respect to the
metric $g_{\mu \nu}$, we obtain the relation $G^{\mu \nu} =
\lambda^{\mu \nu}$; replacing finally this last relation into the
Euler-Lagrange equation with respect to the normal embedding
variations we get the EOM $G^{\mu \nu} \bm{y}_{;\mu\nu} = 0$ in the
bulk, which is formally identical to the equation obtained from the
action principle presented in the previous section. For free
embeddings this EOM implies the Einstein equations, $G^{\mu \nu} =
0.$ However, for $q$-free embeddings,
equivalence with GR can be obtained by supplementing the $10-q$
missing equations by the boundary equations coming from the
arbitrary variation of $\delta W_\mu$ in
eq.(\ref{Var-action-1-degree}) (see previous section).

\subsubsection{The case $n=2$}

Variation of eq.(\ref{actionN}) now reads
\begin{widetext}
\begin{eqnarray}\nonumber
({8\pi G})\, \delta S_2 &=& \int_{\mathcal{M}} \bigg[\delta
\lambda^{\mu \nu \rho \sigma} \left(g_{\mu \nu} -
\bm{y}_{,\mu}\cdot\bm{y}_{,\nu}\right) \left(g_{\rho \sigma} -
\bm{y}_{,\rho}\cdot\bm{y}_{,\sigma}\right)+
\\
\label{Var-action-2-degree} && \qquad\delta g_{\mu \nu} \left(2
\widetilde{\lambda}^{\mu \nu} + \frac{1}{2}g^{\mu
\nu}\widetilde{\lambda}^{\alpha \beta}\left(g_{\alpha \beta}-
\bm{y}_{,\alpha}\cdot\bm{y}_{,\beta}\right)- G^{\mu \nu}[g]\right)
  \bigg]\sqrt{-|g|}
d^4 x
\\
\nonumber &-&4\int_{\partial\mathcal{M}} \delta W_\mu
 \widetilde{\lambda}^{\mu \nu} d
\Sigma_\nu+4\int_{\mathcal{M}} \left[\delta W_\mu
\widetilde{\lambda}_{\quad ;\nu}^{\mu \nu} + \bm{\delta y}_\perp
\cdot \bm{y}_{; \mu \nu}\widetilde{\lambda}^{\mu
\nu}\right]\sqrt{-|g|} d^4 x \,,
\end{eqnarray}
\end{widetext}
where $\widetilde{\lambda}^{\mu \nu} \equiv \lambda^{\mu \nu \rho
\sigma} \left(g_{\rho \sigma} -
\bm{y}_{,\rho}\cdot\bm{y}_{,\sigma}\right) $.

The Euler-Lagrange equations stemming from the first two variations
in eq.(\ref{Var-action-2-degree}) are equivalent to:
\begin{equation}\label{solBDB}
\begin{array}{rcl}
g_{\mu\nu}&=&{\bm{y}}_{,\mu}\cdot{\bm{y}}_{,\nu}\\
G^{\mu \nu}[g] & = & 0\,, \\
\end{array}
\end{equation}
which imply  Einstein dynamics for the embedding variables.
Equations (\ref{solBDB}) also imply that both remaining surface and
bulk variations in eq.(\ref{Var-action-2-degree}) vanish. At this
stage, ${\lambda}^{\mu \nu \alpha \beta}$ remains completely
arbitrary.

It thus seems that, at the level of EOM, freeness is not needed to
recover Einstein equations. However this is misleading because the
existence of propagating gravitational waves equivalent to those of
GR for this theory needs, as we are now going to prove, that the
embedding be $q$-free.

Let us perturb the EOM of the $S_2$ embedding theory by varying the
embedding variables $y^A$, the metric $g_{\mu \nu}$ and the tensor
$\lambda^{\mu\nu\alpha\beta}$. After the perturbation has been
performed to first order, we can use the EOM wherever we want.
Recall that here $G^{\mu \nu}$ depends only on $g_{\mu \nu}\,.$
First note that the variation $\delta \frac{\delta S}{\delta
\lambda^{\mu \nu \alpha \beta}}$ vanishes because of the EOM. The
other variations give, in the bulk:
\begin{equation}\label{deltaEulerLagrG}
   - \delta G^{\mu \nu} + 2
    {\lambda}^{\mu \nu \alpha
    \beta} \delta(g_{\alpha\beta}-\bm{y}_{,\alpha}\cdot \bm{y}_{,\beta})=
    0\,,
\end{equation}
\begin{eqnarray}
\nonumber    0&=&\bm{y}_{,\mu} \left[{\lambda}^{\mu \nu \alpha
\beta} \delta(g_{\alpha\beta}-\bm{y}_{,\alpha}\cdot
    \bm{y}_{,\beta})\right]_{;\nu}\\
\label{deltaEulerLagrYBulk}
    &+& \bm{y}_{; \mu \nu} {\lambda}^{\mu \nu \alpha
    \beta} \delta(g_{\alpha\beta}-\bm{y}_{,\alpha}\cdot \bm{y}_{,\beta})
    \,.
\end{eqnarray}
On the boundary we get:
\begin{equation}\label{deltaEulerLagrYBoundary}
{\mathrm{On \quad{\partial\mathcal{M}}:}} \qquad n_\mu
{\lambda}^{\mu \nu \alpha
    \beta} \delta(g_{\alpha\beta}-\bm{y}_{,\alpha}\cdot \bm{y}_{,\beta})=
    0\,.
\end{equation}
Now, using eq.(\ref{deltaEulerLagrG}),
eqs.(\ref{deltaEulerLagrYBulk}--\ref{deltaEulerLagrYBoundary})
become:
\begin{eqnarray}\label{newdeltaEulerLagrYBulk}
\bm{y}_{,\mu} \left(\delta G^{\mu\nu}\right)_{;\nu} + \bm{y}_{; \mu
\nu} \delta G^{\mu\nu}& = 0, & \quad {\mathrm{on
\,\,\,\,{\mathcal{M}}}}
\,,\\
\label{newdeltaEulerLagrYBoundary} n_\mu \delta G^{\mu\nu}  &=
    0, & \quad {\mathrm{on
\,\,\,\,{\partial\mathcal{M}}}}  \,.
\end{eqnarray}
The first term in eq.(\ref{newdeltaEulerLagrYBulk}) vanishes (see
the discussion after eq.(\ref{deltaRTEulerLagrBulk})). Equations
(\ref{newdeltaEulerLagrYBulk}--\ref{newdeltaEulerLagrYBoundary}) are
thus equivalent to eqs.
(\ref{deltaRTEulerLagrBoundary}--\ref{deltaRTEulerLagrBulk}) and
these are in turn equivalent to Einstein equations when the
embedding is $q$-free.
Note finally that it is necessary to assume that the tensor
${\lambda}^{\mu \nu \alpha \beta}$ is invertible in order to deduce
from eq. (\ref{deltaEulerLagrG}) the embedding equation $\delta
(g_{\mu\nu} - \bm{y}_{,\mu}\cdot \bm{y}_{,\nu}) = 0$, which permits
the identification $g_{\mu \nu}=\bm{y}_{,\mu}\cdot \bm{y}_{,\nu}$ of
the metric in terms of the embedding at the EOM and at the
perturbative level as well; this is what makes possible a
description of gravitational waves in terms of embedding waves.

\section{Vacuum initial value formulation of GR in embedding variables}
\label{sec:initial} In this section we discuss the possibility of
treating the Einstein equations as a dynamical system in terms of
the embedding functions. We present some preliminary results
regarding the initial value formulation of the vacuum Einstein
equations in the embedding approach. Note that, in the usual
formulation, the numerical integration schemes suffer from
instabilities (e.g., pure gauge modes and violation of constraints
\cite{Alc00,Hol04,Sie01}), which destroy their performance in finite
time. In this context, the direct numerical integration of the
equations in the embedding variables is certainly worth developing.

Another motivation for future numerical work is that it could
provide information on some interesting theoretical questions. For
example: is $q$-freeness (or spatially free $q$-freeness) kept as
long as one integrates in time the evolution equations in embedding
variables?

Let us assume for simplicity that the dimension of the host space is $N=10$.

\subsection{The coordinate system and the embedding variables}
\label{subsec:lapse} Let us assume that the host space has only one
time-like coordinate: $\eta_{A B}= {\mathrm{diag}}(1,-1,\ldots,-1),$
and that the space-time is globally hyperbolic. One can then
introduce a coordinate system $(\tau,\lambda^i),$ with $\tau$ a
global time-like coordinate and $g^{\tau\tau}>0.$ Let the initial
embedding, defined by the functions $\bm{y}(\tau=0,\lambda^i)$, be
$6$-free and spatially free with respect to this coordinate system.

Let  $\bm{u} \equiv \frac{\bm{y}^{,\tau}}{\sqrt{g^{\tau\tau}}};$
this vector is timelike and verifes:
\begin{equation}\label{nConstraints}
\begin{array}{rcl}
  \bm{u}\cdot \bm{u} & = & 1 \\
  \bm{u} \cdot \bm{y}_{,i} & = & 0,
\end{array}
\end{equation}
where $F_{,i} \equiv \frac{\partial F}{\partial \lambda^i}.$ Let
$(u^\tau,u^i)$ be the components of $\bm{u}$ in the basis
$\{\bm{y}_{, \tau},\bm{y}_{,i}\},$ i.e., $\bm{u} = u^\tau \bm{y}_{,
\tau} + u^i \bm{y}_{, i}$. One has $u^\tau = (u_\tau)^{-1}=
{\sqrt{g^{\tau\tau}}}$ and $u_i=0.$ Defining now the lapse function
$N(\tau,\lambda^i)$ and the shift vector $N^i(\tau,\lambda^i)$ by
\begin{equation}\label{shift}
 \begin{array}{rcl}
   N & = & {(u^\tau)^{-1}}, \\
   N^i & = & -(u^\tau)^{-1}\,{u^i}, \\
 \end{array}
\end{equation}
one gets the following evolution equation
for ${\bm{y}}$:
\begin{equation}\label{yEOM}
{\bm{y}}_{,\tau} = N \bm{u}+N^i \bm{y}_{, i}\,.
\end{equation}

In terms of the functions $(\bm{y},N,N^i)$, the metric components
read $ g_{ij} = \bm{y}_{,i}\cdot\bm{y}_{,j}\,,\,$ $
  g_{\tau i}  =  - g_{i j} N^j\,,\,$ $ g_{\tau \tau}  =  N^2 + g_{i j} N^i N^j\,.$
The condition for $\bm{y}_{, \tau}$ to be time-like is $g_{\tau\tau}
> 0.$

Our definition of lapse and shift agrees
with the standard one (see Wald
\cite{W84a}). It is well known that these
functions are not dynamical and that fixing
them improperly can cause problems in the
numerical integration of Einstein equations
\cite{Landau}. Because of coordinate
transformation invariance, $4$ components of
$\bm{y}$ can be chosen as arbitrary
functions of $(\tau, \lambda^i)$. The
corresponding $4$ equations in (\ref{yEOM})
then fix $(N,N^i)$ once $\bm{u}$ is known.

In this way, the evolution equations (\ref{yEOM}) allow to propagate
in time the embedding functions $\bm{y}$ provided the vector
$\bm{u}$ is known. As shown in the next section, the choice of
$\bm{u}$ as the relevant variables `conjugate to $\bm{y}$' is the
natural one to deal with Einstein equations.

\subsection{The vacuum Einstein equations}
As usual, the vacuum Einstein equations $G^{\tau j}=0$ and $G^{\tau
\tau}=0$ are $4$ constraints on the initial data; in the present
case the initial data are the values of $(\bm{y}, \bm{u})$ at
$\tau=0$. The first three equations are linear in the derivatives
$\bm{u}_{,j}$, while the last one is algebraic and quadratic in
$\bm{u}:$
\begin{equation}\label{EinsteinConstraints}
\begin{array}{rcl}
\tilde{g}^{j k}\left(\bm{y}_{;i j}\cdot\bm{u}_{,k} - \bm{y}_{;j
k}\cdot\bm{u}_{,i}\right) & = & 0, \quad i = 1,2,3,\\
\tilde{g}^{i j}\tilde{g}^{k l}\left(\bm{y}_{;j k}\cdot\bm{y}_{;i l}-\bm{y}_{;i j}\cdot\bm{y}_{;k l}\right)  & = & 0; \\
\end{array}
\end{equation}
here, $\tilde{g}^{i j}$ is the $3-$inverse of $g_{i j}$ and
$\bm{y}_{;j k} = {\mathbf{S}} \cdot \bm{y}_{,j k}-\bm{u}(
\bm{u}\cdot\bm{y}_{,j k}),$ where the spatial projector
${\mathbf{S}}$ is defined by ${\mathbf{S}}_{A B} = \eta_{A B} -
\tilde{g}^{i j}\,y_{A, i}\, y_{B, j} .$ The projector and the metric
components appearing in eqs.(\ref{EinsteinConstraints}) are clearly
given by the initial data.

Let us define the `acceleration' $\bm{a}$ by
\begin{equation}\label{nEOM}
\frac{\partial\bm{u}}{\partial \tau} \equiv N \bm{a}+{N^i}
\bm{u}_{,i}\,.
\end{equation}

Since $(N,N^i)$ have been fixed by a choice of  coordinates, the time
evolution of $\bm{u}$ will be completely determined if the
acceleration $\bm{a}$ is known. Its normal components are
obtained (when the embedding is spatially free) from the vacuum
Einstein equations $R_{i j}=0$:
\begin{equation}\label{n normal deriv}
\bm{y}_{;i j} \cdot \bm{a} = \bm{u}_{,i}\cdot {\mathbf{S}}\cdot
\bm{u}_{,j} + \tilde{g}^{k l} \left(\bm{y}_{;i k} \cdot \bm{y}_{;j
l} - \bm{y}_{;i j} \cdot \bm{y}_{;k l} \right)
\end{equation}
and its tangent components satisfy:
\begin{equation}\label{nConstraintsDerived}
\begin{array}{rcl}
\bm{u}  \cdot  \bm{a}& = & 0 \\
\bm{y}_{,i}\cdot  \bm{a} & = & -(\ln N)_{,i}\,. \\
\end{array}
\end{equation}
These expressions are obtained from straightforward manipulation
 of the differentiated versions of eqs.(\ref{nConstraints}) and eqs.(\ref{yEOM}).

Equations (\ref{nEOM}--\ref{nConstraintsDerived}) can thus be used to
propagate in time the vector $\bm{u}$.

\section{Conclusion}
\label{sec:conclusion}

\subsection{Summary}

We have revisited the embedding approach on General Relativity which
views the $4$-D, possibly curved, physical space-time as a membrane
floating in a flat host space-time of higher dimension. We have
first introduced two new classes of embeddings, both based on Nash'
notion of freeness. All embeddings in these classes are deformable
and, therefore, allow for a description of gravitational waves;
explicit examples of such embeddings have been given for both
Minkovski space-time and the Schwarzschild black hole. We have also
presented new variational principles which deliver General
Relativity as a field theory for embedding variables. Einstein's
dynamics thus appears as free membrane dynamics in the host space.
We have finally considered the general relativistic initial value
formulation in terms of embedding variables and argued that this new
point of view sheds new light on this particularly difficult issue.

\subsection{Discussion}

This article proposes what is, to the best of our knowledge, the
first consistent embedding approach to non quantum gravitation.
Previous attempts have been marred by essentially two problems. The
first one concerns the possibility of deforming the embedding to
accommodate for gravitational waves. Our approach is the first to use
the notion of freeness introduced by Nash, and variations thereof,
to solve this problem. The second problem is linked with the
possibility of constructing an action principle which would deliver
Einstein's theory in terms of the embedding variables. Previous
attempts \cite{{Reg75},{Des76},{Tap89},{Fra92}} failed in this
respect because the action functionals were used with the wrong
boundary conditions, and also because the considered embeddings were
not free.

We have constructed explicitly free embeddings of several general relativistic
space-times of astrophysical importance (for example, a Schwarzschild black hole).
The approach developed in this article is thus relevant to physics. Whether all general
relativistic space-times can be embedded freely in a flat space-time of higher
dimension remains however an open question.

We would like to end this article by mentioning a few assets offered
by the embedding point of view on GR which have not already been
discussed.

First, the embedding point of view is remarkably modern, offering
obvious links with String theory, M-theory (or F-theory) and
cosmology (see
\cite{randall-1999-83,Mai02,maartens-2004-7,kiritsis-2005-421,davidson-2006-74}
and references therein). For example, embedding variables seem
ideally suited to the semiclassical study of the spontaneous
creation or destruction of universes out of a quantum vacuum
\cite{Lin98}.

More generally, the embedding point of view surely appears as \emph{the} right tool to study problems involving changes in the topology
of space-time. In fact, in our example of flat space-time, the extra
parameters and functions (appearing explicitly in the embedding but
not in the metric) can be chosen in order to change the topology of
space-time from $\mathbb{R}\times \mathbb{R}^3$ to $\mathbb{R}\times
\mathbb{R}^2\times S^1$. In this context, the problem of averaging
statistically the geometry of (classical) space-time has recently
been solved for situations in which the topology of space-time is
fixed \cite{Debb3,Debb}. Does the embedding point of view permit a
more general treatment?

Finally, it is certainly interesting to investigate how field
quantization on the (flat) host space translates on the embedded
space-time.  The free embedding theory of gravity introduced in this
paper, precisely because it deals with $q$-free embeddings, is
suitable for perturbative quantum theory, as opposed to the old
approaches (see, for example, \cite{Fra92}). To cite just a few
interesting questions: does the resulting 4-D quantum theory depend,
at fixed space-time coordinates $x^\mu$, on the choice of the
embedding? We refer the reader to discussions of this topic in the
context of RT theory \cite{{Fra92},Pav94}, and in the context of
embedding theory of induced gravity in \cite{Pav94}. What new
insight does the embedding point of view bring to the Unruh effect
\cite{deser-1998-15}?
 And how does black
hole thermodynamics appear in the embedding point of view? (See
\cite{Birrel,Wald} for an account of these two last topics in the
context of GR).

\appendix

\section {Christoffel
symbols, Covariant derivative, Second Fundamental Form, Normal
Projector, and Curvature tensor in terms of embedding functions}
\label{appendix:covariant} Following Dirac \cite{Dir96} we define
the Christoffel symbols in terms of the embedding functions:
$\Gamma_{\,\,\,\mu \nu}^\alpha = \bm{y}^{,\alpha}\cdot\bm{y}_{, \mu
\nu}\,.$ The covariant derivative is defined as usual in terms of
the Christoffel symbols. For example, the second covariant
derivative of a scalar function $\phi(x^\mu)$ is $\phi_{;\mu \nu} =
\phi_{, \mu \nu} - \phi_{,\alpha} \Gamma_{\,\,\,\mu \nu}^\alpha\,.$
It is interesting to see that, when $\phi$ is replaced by the
embedding functions (which are scalars by definition) we get what is
known as the second fundamental form: $\bm{y}_{;\mu \nu} = \bm{y}_{,
\mu \nu} - \bm{y}_{,\alpha} \Gamma_{\,\,\,\mu \nu}^\alpha =
\bm{y}_{, \mu \nu} - \bm{y}_{,\alpha}\,
(\bm{y}^{,\alpha}\cdot\bm{y}_{, \mu \nu}) = {\mathbf{N}}\cdot
\bm{y}_{, \mu \nu} \,,$ where ${\mathbf{N}}^{A B} = \eta^{A B} -
(y^A)_{,\alpha}(y^B)^{,\alpha}$ is the normal projector, whose
kernel is by definition the tangent space ${{T}_p}( {\mathcal{M}}),$
generated by the tangent vectors $\bm{y}_{, \mu}(x^\nu), \,\mu = 0,
\ldots, 3.$ Then, the second fundamental form is a set of vectors in
the normal space ($\bm{y}_{;\mu \nu} \in {{N}_p}( {\mathcal{M}}),
\quad \mu,\nu = 0, \ldots 3$) with $\bm{y}_{;\mu \nu} \cdot
\bm{y}_{,\alpha} =0.$

Finally we compute the Riemann and Einstein tensors in terms of the
embedding functions. The Riemann curvature tensor depends on second
(partial) derivatives of the metric tensor, which itself depends on
first derivatives of the embedding functions. Therefore one could
naively expects the curvature tensor to depend on third order
derivatives of the embedding functions. However, all terms
containing third order derivatives vanish. The Riemann tensor thus
depends on second derivatives only and reads $R_{\mu \nu \alpha
\beta} = \bm{y}_{;\mu \alpha} \cdot \bm{y}_{;\nu \beta} -
\bm{y}_{;\nu \alpha} \cdot \bm{y}_{;\mu \beta}.$ The Einstein tensor
is similarly written $G_{\alpha\beta} =
g^{\kappa\mu}\left(\bm{y}_{;\kappa\nu}\cdot
\bm{y}_{;\lambda\mu}-\bm{y}_{;\kappa\mu}\cdot
\bm{y}_{;\lambda\nu}\right)\left[\delta_\alpha^\lambda\,\delta_\beta^\nu
- \frac{1}{2} g_{\alpha\beta}g^{\lambda\nu}\right].$

Geometrically, it is natural that only second covariant derivatives
of embedding functions appear in these tensors: the curvature radii
of the embedded manifold along the principal axes depend essentially
on the \textit{normal} components of the matrix of second
derivatives of the embedding functions (the second fundamental
form).


\end{document}